\documentstyle[multicol,aps,psfig]{revtex}
\renewcommand{\narrowtext}{\begin{multicols}{2}
\global\columnwidth20.5pc\noindent}
\renewcommand{\widetext}{\end{multicols}
\global\columnwidth42.5pc}
\begin{document}
\draft
\preprint{\today}
\title{
Two Kinds of the Coexistent States 
in One-Dimensional Quarter
-Filled Systems under Magnetic Fields 
}

\author{Keita Kishigi and Yasumasa Hasegawa}
\address
{Faculty of Science, Himeji Institute of Technology,
 Ako, Hyogo 678-1297, Japan}
\date{\today}
\maketitle
\begin{abstract}
The coexistent state of the spin density wave (SDW) and
the charge density wave (CDW) in the one-dimensional
quarter-filled system and with the Coulomb interaction
up to the next-nearest sites 
under magnetic fields is studied.
It is found that, 
in the coexistent state of 
$2k_{\rm F}$-SDW and $2k_{\rm F}$-CDW, 
the charge order is suppressed 
and $2k_{\rm F}$-SDW changes to $2k_{\rm F}$-SDW 
having the different alignment of spin under high magnetic fields, 
whereas, in the coexistent state of $2k_{\rm F}$-SDW and $4k_{\rm F}$-CDW, 
$4k_{\rm F}$-CDW still remains and 
$2k_{\rm F}$-SDW is suppressed. 
The critical temperature of the charge order is 
higher than that of the spin order when the inter-site 
Coulomb interaction is strong. 
These will be observed in experiments such as the X-ray 
scattering measurement and NMR. 
\end{abstract}
\narrowtext
\section{Introduction}

Organic conductors
such as (TMTSF)$_2$$X$ and (TMTTF)$_2$$X$
($X$=ClO$_4$, PF$_6$, AsF$_6$, ReO$_4$, Br, SCN, etc.)
are the quasi-one dimensional quarter-filled systems and
exhibit various kinds of ground states,
for example, spin-Peierls, 
spin density wave (SDW) and 
superconductivity.\cite{review,jerome}
In (TMTSF)$_2$PF$_6$, the incommensurate SDW occurs at $T=12$ K,
where the wave vector is determined by NMR
experiments\cite{takahashi,delrieu} as (0.5, 0.24, 0.06). 
Pouget and Ravy observed 
the coexistence of $2k_{\rm F}$-SDW and
$2k_{\rm F}$-charge density wave (CDW) 
by the X-ray scattering measurement,\cite{pouget} 
where $k_{\rm F}=\pi /4a$ 
is Fermi wave number and
$a$ is the lattice constant.
The critical temperature ($T_{\rm CDW}$) at which 
$2k_{\rm F}$-CDW is observed is the same temperature as $T_{\rm SDW}$.

On the other hand, it is found that SDW in 
(TMTTF)$_2$$X$ ($X$=Br and SCN)
has the commensurate wave vector, 
(0.5, 0.25,0), as observed in the measurements of
$^{13}$C-NMR\cite{barthel} and $^1$H-NMR\cite{nakamura}.
From the angle depenence of satelite peak positions of
$^1$H-NMR\cite{nakamura,nakamura2},
the alignment of the spin moment along the conductive axis (a-axis)
is obtained to be ($\uparrow,0,\downarrow,0$), 
where the arrow means the spin moment. 
In (TMTTF)$_2$Br, $4k_{\rm F}$-CDW accompanied by $2k_{\rm F}$-SDW
is found in X-ray scattering measurments, where 
$T_{\rm CDW}\sim 100$ K and $T_{\rm SDW}\sim 13$ K.\cite{pouget} 

In the one-dimensional quarter-filled systems, 
the CDW-SDW coexistent state 
is shown to be caused by 
the interplay between the on-site Coulomb interaction ($U$) and the
inter-site Coulomb interaction ($V$).
\cite{seofukuyama,nobuko,nobuko2,Mazumdar,yoshioka} 
The inter-site Coulomb interaction plays 
important role for the charge order. 
Mila\cite{mila} has estimated $U/t\sim5$ and $V/t\sim2$, 
where 
$t$ is a transfer integral. 
Seo and Fukuyama\cite{seofukuyama} showed that
the ground state becomes the coexistent state of
$2k_{\rm F}$-SDW and $4k_{\rm F}$-CDW in the one-dimensional
extended Hubbard model with $V$. 
It is found by Kobayashi et al.\cite{nobuko,nobuko2} and 
Mazumdar et al.\cite{Mazumdar} that
the coexistent state of $2k_{\rm F}$-SDW and $2k_{\rm F}$-CDW
is stabilized 
when the next nearest neighbor Coulomb interaction ($V_2$)
and the dimerization of the
energy band are considered. 
Thus, two kinds of the 
coexistent states ($2k_{\rm F}$-SDW-$2k_{\rm F}$-CDW and 
$2k_{\rm F}$-SDW-$4k_{\rm F}$-CDW), 
which are observed 
by the X-ray scattering mesauerments 
in (TMTSF)$_2$PF$_6$ and (TMTTF)$_2$Br, 
can be explained by 
Seo and Fukuyama\cite{seofukuyama} and 
Kobayashi et al.\cite{nobuko,nobuko2} and 
Mazumdar et al.\cite{Mazumdar}, 
respectively.

Recently, we have studied the effects of the magnetic 
field ($H$) on 
the coexistent state of $2k_{\rm F}$-SDW and $4k_{\rm F}$-CDW
for both cases of strong and weak coupling 
of the correlation between electrons, where 
the one-dimensional quarter-filled extended
Hubbard model with $V$ is used.\cite{kishigi}
It is found that although the spin order is suppressed at
high fields, the charge order still remains
in the strongly coupling systems ($U/t\sim 5$). 
When the coupling is small ($U/t\sim 1.5$),
both orderings of $2k_{\rm F}$-SDW and $4k_{\rm F}$-CDW
disappear at the critical magnetic field, which is the 
same as the Pauli paramagnetic field in the spin-singlet 
superconductivity.\cite{pauli,pauli2,Mckenzie}

In this paper, we examine how
the coexistent state of $2k_{\rm F}$-SDW and
$2k_{\rm F}$-CDW and the one of $2k_{\rm F}$-SDW and
$4k_{\rm F}$-CDW are affected by magnetic fields and temperatures. 
We use the one-dimensional extened Hubbard model
with $V$ and 
$V_2$, where 
the dimerization is neglected for simplicity, 
because the coexistent states are stabilized without 
the dimerization. 
We use parameters, $U/t=5.0$, which is indicated 
in (TMTTF)$_2$$X$
((TMTSF)$_2$$X$).\cite{band,band2,band3,band4} 
In most cases we set 
$V/U\leq 1$ and $V_2<V$. 
We study the case when the anisotorpy of the 
spin is strong, because it is found that 
the easy axis of the spin in 
the quasi-one-dimensional organic conductors 
is $b$-axis.\cite{spin} 
In order to clarify the ground state under 
fields at finite temperatures, we show the $V$-$H$, $V_2$-$H$, 
$V$-$T$ and $V_2$-$T$ 
phase diagrams, which enables to discuss 
the effect of pressure and the ordering temperatures of 
the SDW and CDW. 
It is expected that the Hubbard model with large 
$U/t\sim5$ is qualitatively understood by the 
Ising model with the perpendicular field, as we will 
show below.

The $H$-dependence of the 
antiferromagnetic state in one-dimensional Ising model has been
studied in the mean field approximation, 
where the componet of the spin 
along the $c$-axis is considered.\cite{ising,ising2} 
When the magnetic fields is applied perpendicular 
to the spin moment, 
\begin{eqnarray}
S_z(H,j)/S_z(0,j)=\sqrt{1-(H/H_x^0)^2},
\end{eqnarray}
where
$S_z(H,j)$ is the amplitude of the spin moment 
at $j$ site at $H=0$ and $H_x^0$ is the critical field at which the
ordering of the antiferromagnetic state disappear 
($H_x^0\propto J$, where $J$ is the coupling constant).\cite{ising,ising2} 



\section{Formulation}

We study the one-dimensional extended
Hubbard model,

\begin{eqnarray}
\hat{\cal H}&=&\hat{\cal K}+\hat{\cal U}+\hat{\cal V}+\hat{\cal V}_2, \\
\hat{\cal K}&=&-t\sum_{i,\sigma}(c^{\dagger}_{i,\sigma} c_{i+1,\sigma}+
h.c.) \nonumber \\
&&+\frac{\mu_{\rm B}g}{2}\sum_{i, \sigma,\sigma^{\prime}}c^{\dagger}_{i,\sigma}
(H_{x})_{\sigma\sigma^{\prime}}c_{i,\sigma^{\prime}},\\
\hat{\cal U}&=&U\sum_{i}n_{i, \uparrow}n_{i, \downarrow}, \\
\hat{\cal
V}&=&V\sum_{i,\sigma,\sigma^{\prime}}n_{i,\sigma}n_{i+1,\sigma^{\prime
}}, \\
\hat{\cal 
V}_2&=&V_2\sum_{i,\sigma,\sigma^{\prime}}n_{i,\sigma}n_{i+2,\sigma^{\prime}},
\end{eqnarray}
where $\mu_{\rm B}=e\hbar/2m_{0}c$ is the Bohr magneton and
$c^{\dagger}_{i,\sigma}$ is the creation
operator of $\sigma$ spin electron at $i$ site,
$n_{i,\sigma}$ is the number operator, $g=2$,
$i=1,\cdots,N_{\rm S}$, $N_{\rm S}$ is the
number of the total sites and $\sigma =\uparrow$ and $\downarrow$. 
The notation in this paper follows Seo and Fukuyama.\cite{seofukuyama} 
The magnetic field is
applied to the x-axis and 
$(H_{x})_{\sigma\sigma^{\prime}}$
$=H(\hat{\sigma}_{x})_{\sigma\sigma^{\prime}}$, 
where 
$H$ is the strength of the magnetic field and
$\hat{\sigma}_{x}$ is Pauli spin matrix.
We consider the quarter-filled case.

The interaction terms, $\hat{\cal U}$,
$\hat{\cal V}$ and $\hat{\cal V}_2$ are treated in the mean field
approximation as 

\begin{eqnarray}
\hat{\cal U}^{\rm M}&=&\sum_{k_{x}}
\sum_{Q}\{
\rho_{\uparrow\uparrow}(Q,T)
C^{\dagger}(k_{x},
\downarrow)
C(k_{x}-Q,\downarrow)  \nonumber \\
&+&\rho^{*}_{\downarrow\downarrow}(Q,T)
C^{\dagger}(k_{x}-Q,
\uparrow)
C(k_{x},\uparrow)\} \\
\hat{\cal V}^{\rm M}&=&(\frac{V}{U})\sum_{k_{x},
\sigma,\sigma^{\prime}}\sum_{Q}e^{-iQa}\{
\rho_{\sigma\sigma}(Q,T)
C^{\dagger}(k_{x},
\sigma^{\prime})
C(k_{x}-Q,
\sigma^{\prime})  \nonumber \\
&+&\rho_{\sigma^{\prime}\sigma^{\prime}}^{*}(Q,T)
C^{\dagger}(k_{x},
\sigma)
C(k_{x}-Q,
\sigma)\} \\
{\hat{\cal V}_2}^{\rm M}&=&(\frac{V_2}{U})\sum_{k_{x},
\sigma,\sigma^{\prime}}\sum_{Q}e^{-2iQa}\{
\rho_{\sigma\sigma}(Q,T)
C^{\dagger}(k_{x},
\sigma^{\prime})
C(k_{x}-Q,
\sigma^{\prime})  \nonumber \\
&+&\rho_{\sigma^{\prime}\sigma^{\prime}}^{*}(Q,T)
C^{\dagger}(k_{x},
\sigma)
C(k_{x}-Q,
\sigma)\} 
\end{eqnarray}
where $I=U/N_{\rm S}$.
The self-consistent equation for the order parameter 
$\rho_{\sigma\sigma}(Q,T)$ at finite temperature, $T$, 
is given by
\begin{eqnarray}
\rho_{\sigma\sigma}(Q,T)&=&I\sum_{k_{x}}
<C^{\dagger}(k_{x},
\sigma)
C(k_{x}-Q,
\sigma)>.
\end{eqnarray}
In eqs. (7)-(10), we take 
the order parameters, $\rho_{\sigma\sigma}(Q,T)$, 
only between electrons with same spins. 
We do not consider the case of
the mean field,
${\rho}_{\sigma
\bar{\sigma}}(Q,T)=I\sum_{k_{x}}<C^{\dagger}(k_{x},\sigma)C(k_
{x}-Q,\bar{\sigma})>$ with $\sigma\neq\bar{\sigma}$. 
We neglect the effect of the spin tilting to 
the $x-y$ plane by setting ${\rho}_{\sigma \bar{\sigma}}(Q,T)=0$ 
in this mean field approximation. 
This simplification corresponds to the assumption that
the rotational 
symmetry in the spin-space is broken and that the $z$-axis 
is 
the easy axis 
although the spin in this Hubbard model is isotropic. 
The magnetic field is applied perpendicular to 
the easy axis when $H\neq 0$.


In the Fulde-Ferrel-Larkin-Ovchinnikov state in the
superconducutivity\cite{fulde,fulde2}, 
the total moment of the Cooper 
pair is changed by the magnetic field. 
Similar situation may occur in SDW and CDW, i.e., the 
wave vector $Q$ may be changed by the magnetic field. 
However, 
such state will be stabilized only in the low temperature 
and strong magnetic field region and will not 
change the essential feature studied in this paper. 
Therefore, we take the possible wave vectors of the 
order parameters as 
$Q=Q_0,2Q_0,3Q_0$ and $4Q_0$ (equivalent to 0), where 
$Q_0=2k_{\rm F}$.

We evaluate the Helmholtz free energy
by using the eigenvalue $\epsilon_{j}$ and the unitary matrix
$U_{{\bf k}\sigma,j}$ obtained by diagonalizing the mean field
Hamiltonian 
$\hat{\cal K}+\hat{\cal U}^{\rm M}+\hat{\cal V}^{\rm M}+\hat{\cal
V}_{2}^{\rm M}$, where 
index $j$ includes the degree of the spin freedom.
We determine the chemical potential $\mu (\rho_{\sigma\sigma}(Q,T))$ 
from $N=1/4$, where 

\begin{eqnarray}
N &=& \sum_{k_{x},\sigma}
<C^{\dagger}(k_x,\sigma)
C(k_x,\sigma)> \nonumber \\
&=& \frac{1}{2N_{\rm s}} \sum_{j=1}^{2N_{\rm s}}
\left[ \exp\left(\frac{\epsilon_{j}-\mu}{T}\right) +1
\right]^{-1}.
\end{eqnarray}

The self-consistency condition (eq. (10)) 
is given by 

\begin{eqnarray}
\rho_{\sigma\sigma}(Q,T)=I\sum_{k_x}
\sum_{j}
U^{*}_{(k_{x}
, \sigma), j}
U_{(k_{x}-Q,
\sigma),  j}
f(\epsilon_{j}),
\end{eqnarray}
where $f(\epsilon_{j})$ is the Fermi function.

We obtain the free energy per
site 
\begin{eqnarray}
F(\rho_{\sigma\sigma}(Q,T))&=&2\mu n -\frac{T}{N_{\rm s}} \sum_{j=1}^{2N_{\rm s}} \log
\left\{
           \exp\left(\frac{\mu-\epsilon_{j}}{T} \right) +1
\right\} \nonumber \\
&-&\frac{1}{U}\sum_{Q}\rho_{\uparrow\uparrow}(Q,T)\rho^{*}_{\downarrow\downarrow}(Q,T), \nonumber
\\
&-&\frac{V}{U^2}\sum_{Q,\sigma,\sigma^{\prime}}e^{-iQa}\rho_{\sigma\sigma}(Q,T)
\rho_{\sigma^{\prime}\sigma^{\prime}}^{*}(Q,T) \nonumber \\
&-&\frac{V_2}{U^2}\sum_{Q,\sigma,\sigma^{\prime}}e^{-2iQa}\rho_{\sigma\sigma}(Q,T)
\rho_{\sigma^{\prime}\sigma^{\prime}}^{*}(Q,T),
\end{eqnarray}
At $T=0$ 
it reduces to 
the ground state energy per site
\begin{eqnarray}
E(H,\rho_{\sigma\sigma}(Q,0))=&&\frac{1}{N_s}\sum^{N_{s}/2}_{j=1}\epsilon_{j}
-\frac{1}{U}\sum_{Q}\rho_{\uparrow\uparrow}(Q,0)\rho^{*}_{\downarrow\downarrow}(Q,0), \nonumber
\\
&-&\frac{V}{U^2}\sum_{Q,\sigma,\sigma^{\prime}}e^{-iQa}\rho_{\sigma\sigma}(Q,0)
\rho_{\sigma^{\prime}\sigma^{\prime}}^{*}(Q,0) \nonumber \\
&-&\frac{V_2}{U^2}\sum_{Q,\sigma,\sigma^{\prime}}e^{-2iQa}\rho_{\sigma\sigma}(Q,0)
\rho_{\sigma^{\prime}\sigma^{\prime}}^{*}(Q,0).
\end{eqnarray}

The electron density at the $j$ site, $n(j,T)$, 
its deviation from the mean value (1/2), 
$\delta(j,T)$, and the
spin moment at $j$ site, $S_z(j,T)$, are given by
\begin{eqnarray}
n(j,T)=\frac{1}{U}\sum_{Q,\sigma}\rho_{\sigma\sigma}(Q,T)e^{iQja}=\frac{1}{2}+\delta(j,T),
\end{eqnarray}
and 
\begin{eqnarray}
S_z(j,T)=\frac{1}{2U}\sum_{Q}(\rho_{\uparrow\uparrow}(Q,T)
-\rho_{\downarrow\downarrow}(Q,T))e^{iQja}.
\end{eqnarray}
When $T=0$, $n(j)\equiv n(j,0)$, $S_z(j)\equiv S_z(j,0)$ and 
$\delta(j)\equiv\delta(j,0)$.

\begin{figure}
\mbox{\psfig{figure=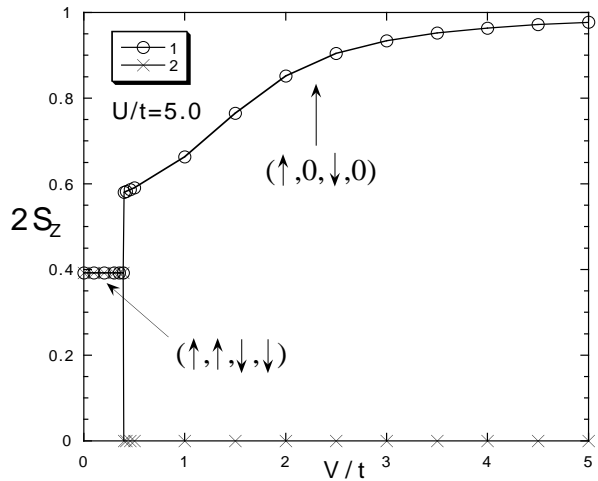,width=80mm,angle=0}}
\vskip 5mm
\caption{2S$_z$(1) and 2S$_z$(2) as a function of $V$ at $H=0$
}
\label{fig1a}
\end{figure}
\vskip 3mm

\begin{figure}
\mbox{\psfig{figure=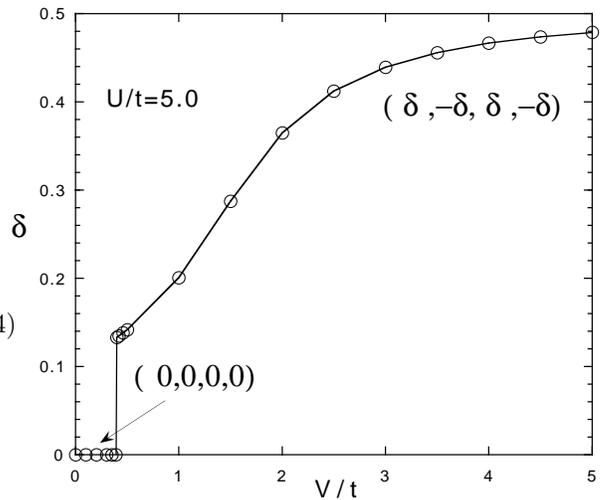,width=80mm,angle=0}}
\vskip 5mm
\caption{$\delta$ as a function of $V$ at $H=0$
}
\label{fig1b}
\end{figure}
\vskip 3mm

\section{Results and Discussions}
\subsection{2k$_{\rm F}$-SDW and 4k$_{\rm F}$-CDW}
In this subsection, we take $V_2=0$ and $T=0$.
It is known that the coexistent state of 
2$k_{\rm F}$-SDW and 4$k_{\rm F}$-CDW are 
induced by $U$ and $V$.\cite{seofukuyama}.
Figs. 1 and 2 are $S_z$ and $\delta$
as a function of $V/t$
at $U/t=5.0$ and $H=0$.
For $0\leq V\leq 0.39$,
the antiferromagnetic order 
(($\uparrow$,$\uparrow$,$\downarrow$,$\downarrow$), i.e.,
$S_z(1)=S_z(2)=-S_z(3)=-S_z(4)$)
is stabilized and there is no charge 
order ($\delta(1)=\delta(2)=\delta(3)=\delta(4)=0$). 
The spin order of
($\uparrow$,$\uparrow$,$\downarrow$,$\downarrow$) has
the wave vector of $2k_{\rm F}$.
For $V/t>0.39$, 
the spin order becomes
($\uparrow$,0,$\downarrow$,0)
($S_z(1)=-S_z(3)$, $S_z(2)=S_z(4)=0$)
and the charge order ($\delta$,$-\delta$,$\delta$,$-\delta$) coexists, 
where 
$\delta(1)=\delta(3)=\delta$ and $\delta(2)=\delta(4)=-\delta$. 
These orders, ($\uparrow$,0,$\downarrow$,0) and
($\delta$,$-\delta$,$\delta$,$-\delta$), 
mean $2k_{\rm F}$-SDW and $4k_{\rm F}$-CDW, respectively.
These results were obtained by Seo and Fukuyama.\cite{seofukuyama}

We show
$S_z$ and $\delta$ for $U/t=5.0$ and $V/t=0\sim 5.0$ as a function 
of perpendicular field ($h_x\equiv \mu_{\rm B}H/t$) in Figs. 3 and 4. 
The antiferromagnetic state is gradually suppressed up to
the critical field 
($h_x^{c}$=1.3, 1.3, 1.5, 1.8 and 2.4 for $V/t$=0, 1.0, 2.0, 2.5 and 5.0, 
respectively), 
as shown in Fig. 3, 
where $S_z(1)=-S_z(3)$ and $S_z(2)=S_z(4)=0$. 
Upon increasing $h_x$ 
the charge order decreases and becomes zero when $V/t=1.0$, and 
the charge order becomes 
constant for $h_x>h_x^{c}$ 
when $V/t\geq 1.5$, 
as shown in Fig. 4, where 
$\delta(1)=\delta(3)=\delta$ and $\delta(2)=\delta(4)=-\delta$. 
The $h_x$-dependence of the amplitude of 
$S_z$ is almost the same as 
eq. (1) when we set 
$H_x^0$ as $H_x^0=th_x^c/\mu_{\rm B}$,
which is shown by solid lines
in Fig. 3. 
From Figs. 3 and 4, we make the $V$-$h_x$ 
phase diagram, as shown in Fig. 5, 
where (0,0,0,0) means that the spin and 
charge orders do not 
exist. The dotted lines show the second order 
transition lines and the solid line show the 
first order 
transition lines. 
In the region of large $h_x$ and $V$, 
the state of $4k_{\rm F}$-CDW, 
($\delta$,$-\delta$,$\delta$,$-\delta$), is 
stabilized.

The $h_x$
-dependence of the spin order 
can be understood 
by the mean field solutions of 
the Ising 
model mentioned 
in the introduction.\cite{ising,ising2}

  
\begin{figure}
\mbox{\psfig{figure=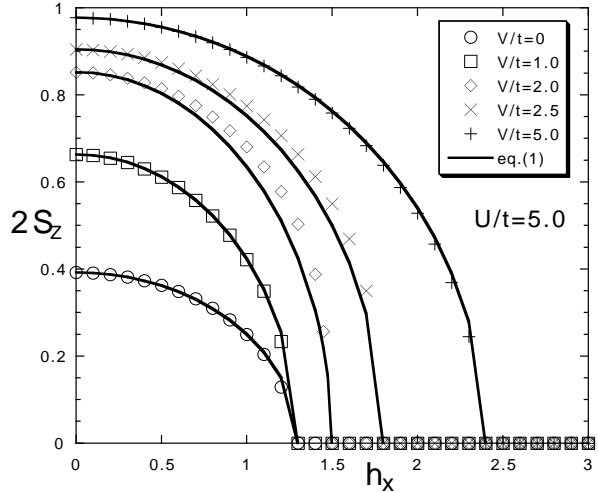,width=80mm,angle=0}}
\vskip 5mm
\caption{
2S$_z$(1) as a function of 
$h_x$. 
The solid line is given by
eq. (1).
}
\label{fig3}
\end{figure}
\vskip 3mm

\begin{figure}
\mbox{\psfig{figure=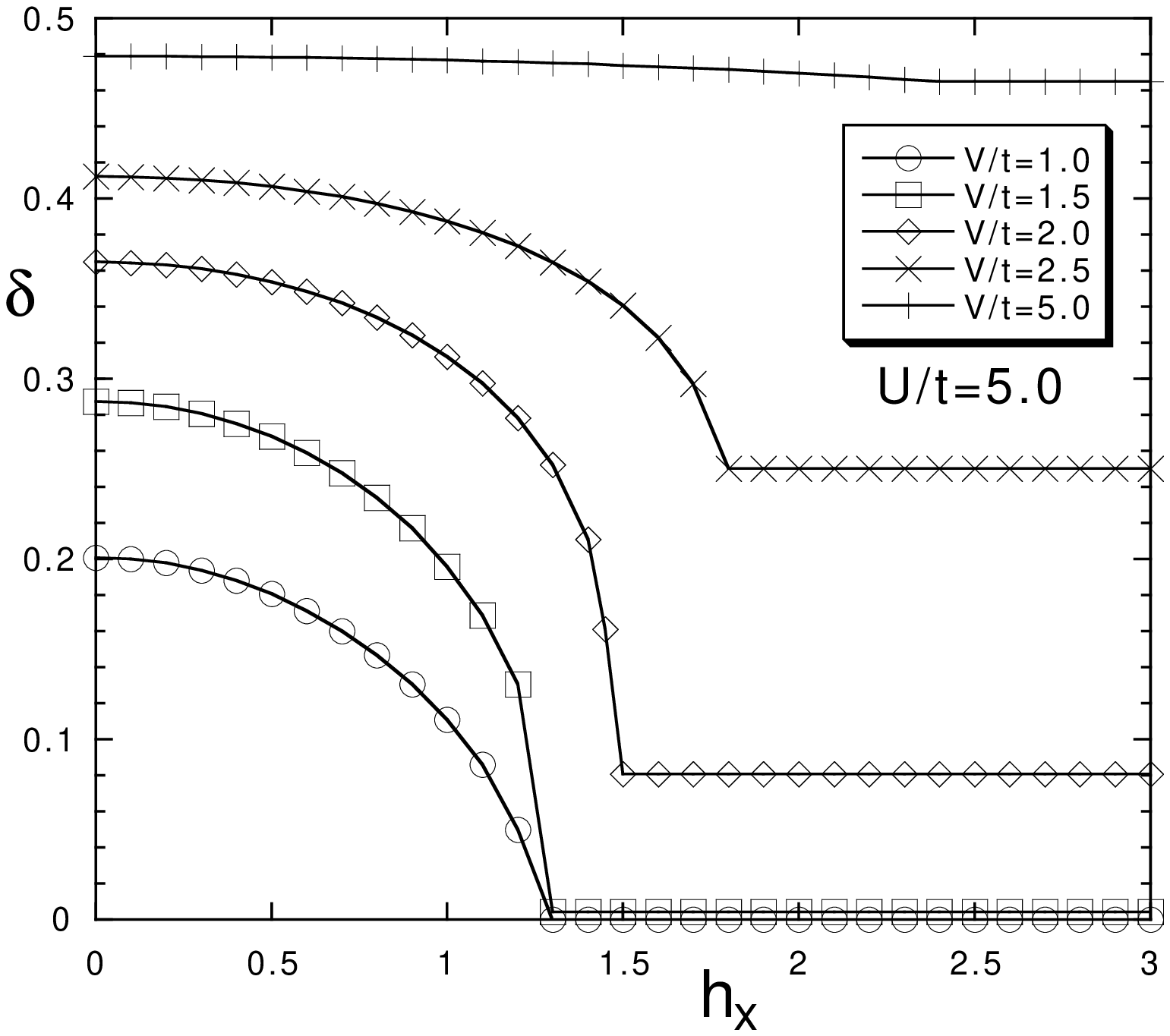,width=80mm,angle=0}}
\vskip 5mm
\caption{
$\delta$ as a function of $h_x$.
}
\label{fig4}
\end{figure}
\vskip 3mm

\begin{figure}
\mbox{\psfig{figure=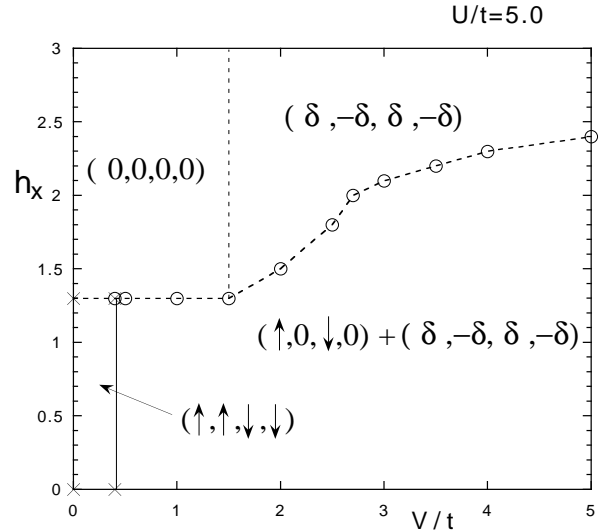,width=80mm,angle=0}}
\vskip 5mm
\caption{
$V-h_x$ Phase diagram.
}
\label{hx-v}
\end{figure}
\vskip 3mm





\subsection{2k$_{\rm F}$-SDW and 2k$_{\rm F}$-CDW}
Here, $V_2$ is introduced, which favors the coexistent state 
of 2$k_{\rm F}$-SDW and 2$k_{\rm F}$-CDW.\cite{nobuko2,Mazumdar}
We search the most stable solution at $U/t=5.0$, $V/t=2.0$, 
$T=0$ and $H=0$ by 
changing $V_2$. 
These $V_2$-dependences of the spin density and 
charge density are shown in Figs. 6 and 7. 
For $0\leq V_2/t\leq 1.4$, 
the coexistent state of 
2$k_{\rm F}$-SDW and 4$k_{\rm F}$-CDW is stable, while 
the state changes to 
the coexistent state 
of 2$k_{\rm F}$-SDW and 2$k_{\rm F}$-CDW 
(($_{\uparrow}$,$\uparrow$,$\downarrow$,$_{\downarrow}$), i.e.,
$S_z(1)=-S_z(4)<S_z(2)=-S_z(3)$) 
and (($-\delta$,$\delta$,$\delta$,$-\delta$), 
$\delta(1)=\delta(4)=-\delta$ and $\delta(2)=\delta(3)=\delta$) 
for $V_2/t>1.4$. 
This transition is a first order transition. 

The coexistent state of 2$k_{\rm F}$-SDW and 2$k_{\rm F}$-CDW 
obtained by 
Kobayashi et al.\cite{nobuko2} 
is (${\uparrow}$,$\uparrow$,$\downarrow$,${\downarrow}$) 
and ($-\delta$, $\delta$, $\delta$, $-\delta$), 
which is different from 
the coexistent state of 2$k_{\rm F}$-SDW and 2$k_{\rm F}$-CDW 
(($_{\uparrow}$,$\uparrow$,$\downarrow$,$_{\downarrow}$) 
and ($-\delta$, $\delta$, $\delta$, $-\delta$)). 
Since they limited 
the freedom of the order parameters, 
they could not find this 
coexistent state of 2$k_{\rm F}$-SDW and 2$k_{\rm F}$-CDW 
(($_{\uparrow}$,$\uparrow$,$\downarrow$,$_{\downarrow}$) and 
($-\delta$, $\delta$, $\delta$, $-\delta$)). 
They also indicated that 
the coexistent state of 2$k_{\rm F}$-SDW and 2$k_{\rm F}$-CDW 
((${\uparrow}$,$\uparrow$,$\downarrow$,${\downarrow}$) and 
($-\delta$, $\delta$, $\delta$, $-\delta$)) is 
suppressed when the dimerization is not included.
In fact, we could not find 
the coexistent state of 
2$k_{\rm F}$-SDW and 2$k_{\rm F}$-CDW of 
(${\uparrow}$,$\uparrow$,$\downarrow$,${\downarrow}$) 
and ($-\delta$, $\delta$, $\delta$, $-\delta$) 
in the parameter region, ($U/t=5.0$, $V/t=2.0$ and $V_2/t<2.0$). 
However, as we show here, 
the coexistent state of 
2$k_{\rm F}$-SDW and 2$k_{\rm F}$-CDW 
(($_{\uparrow}$,$\uparrow$,$\downarrow$,$_{\downarrow}$) 
and ($-\delta$, $\delta$, $\delta$, $-\delta$)) 
is stabilized even in 
the absence of the 
dimerization.  

After we finished this study, we know the very recent study 
at $T=0$\cite{tomio} and $T\neq0$\cite{tomio2} 
in the absence of the magnetic field by Tomio and 
Suzumura. They calculate the mean field solution 
using the extended Hubbard model with $V$ and $V_2$. 
They show that the state of 2$k_{\rm F}$-SDW and 2$k_{\rm F}$-CDW 
(($_{\uparrow}$,$\uparrow$,$\downarrow$,$_{\downarrow}$) 
and ($-\delta$, $\delta$, $\delta$, $-\delta$)) 
coexists even if the dimerization is not included, too.\cite{tomio}

\begin{figure}
\mbox{\psfig{figure=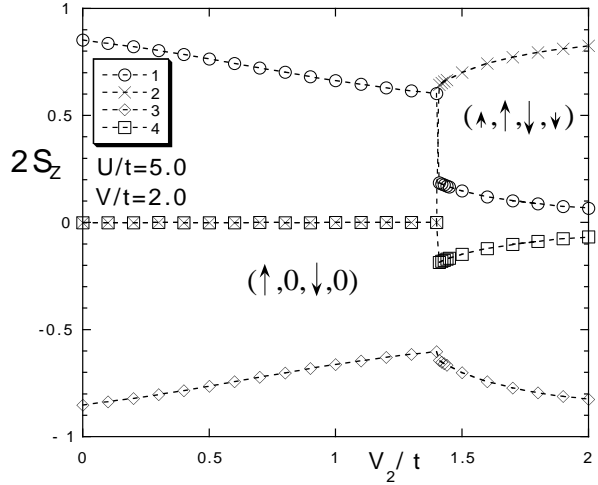,width=80mm,angle=0}}
\vskip 5mm
\caption{
$2S_z(j)$ $(j=1\cdots 4)$ as a function of $V_2/t$ at $H=0$.
}
\label{fig7}
\end{figure}
\vskip 3mm

\begin{figure}
\mbox{\psfig{figure=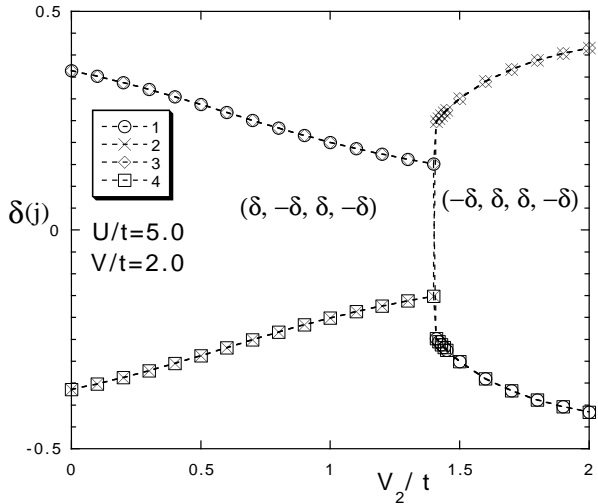,width=80mm,angle=0}}
\vskip 5mm
\caption{
$\delta(j)$ $(j=1\cdots 4)$ as a function of $V_2/t$ at $H=0$.
}
\label{fig8}
\end{figure}
\vskip 3mm

We analyze the coexistent state of $2k_{\rm F}$-SDW and $2k_{\rm F}$-CDW
under magnetic fields at $0\leq V_2/t\leq 2.0$. 
In the case of the perpendicular field ($h_x$), 
the state of 
($_{\uparrow}$,$\uparrow$,$\downarrow$,$_{\downarrow}$) and 
($-\delta$,$\delta$,$\delta$,$-\delta$) becomes the 
one of ($\uparrow$,$\uparrow$,$\downarrow$,$\downarrow$) and
(0,0,0,0) ($\delta(1)=\delta(2)=\delta(3)=\delta(4)=0$) 
at $0.75<h_x<1.25$, which is 
shown in Figs. 8 and 9.
We can see that 
the coexistent state of $2k_{\rm F}$-SDW and $2k_{\rm F}$-CDW 
changes to the state of $2k_{\rm F}$-SDW. 
This transitions is 
the second order phase transition, since 
$|S_z(1)|$ and $|S_z(3)|$ ($|S_z(2)|$ and $|S_z(4)|$) increase
(decrease) gradually, and 
($_{\uparrow}$,$\uparrow$,$\downarrow$,$_{\downarrow}$) 
changes to 
($\uparrow$,$\uparrow$,$\downarrow$,$\downarrow$) 
smoothly. 
When $h_x>1.25$, 
the $2k_{\rm F}$-SDW state disappears. 
For $0\leq V_2/t\leq 1.4$, as $h_x$ increases, 
the spin density 
in the coexistent state of $2k_{\rm F}$-SDW and 
$4k_{\rm F}$-CDW decrease and becomes zero at 
$h_x^c=1.3, 1.3, 1.3$ and 1.5 for $V_2/t=1.4, 1.0, 0.5$ and 0, 
as shown in Fig. 10, where 
$S_z(1)=-S_z(3)$ and $S_z(2)=S_z(4)=0$. 
When we set $H_x^0$ as $H_x^0=th_x^c/\mu_{\rm B}$, 
the $h_x$-dependence of $S_z$ is in agreement with 
eq. (1), which is shown by solid line in Fig. 10.
The order of $4k_{\rm F}$-CDW becomes nonzero at the higher field 
for $V_2/t<0.5$, as shown in Fig. 11, where 
$\delta(1)=\delta(3)=\delta$ and $\delta(2)=\delta(4)=-\delta$. 
We show the $V_2$-$h_x$ phase in Fig. 12, where 
the solid line is for the first order transition and 
the dotted lines are for the second order transitions.



\begin{figure}
\mbox{\psfig{figure=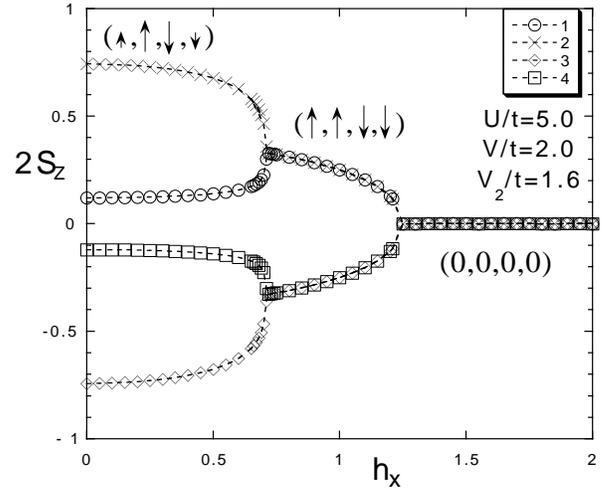,width=80mm,angle=0}}
\vskip 5mm
\caption{
$2S_z(j)$ $(j=1\cdots 4)$ as a function of $h_x$.
}
\label{fig9}
\end{figure}
\vskip 3mm

\begin{figure}
\mbox{\psfig{figure=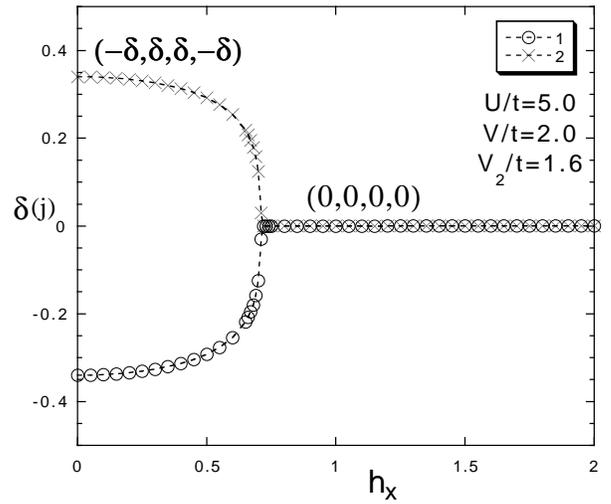,width=80mm,angle=0}}
\vskip 5mm
\caption{
$\delta(1)$ and $\delta(2)$ as a function of $h_x$.
}
\label{fig10}
\end{figure}
\vskip 3mm

\begin{figure}
\mbox{\psfig{figure=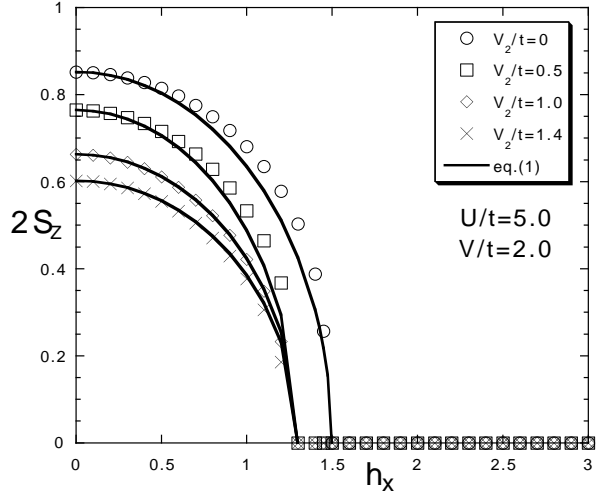,width=80mm,angle=0}}
\vskip 5mm
\caption{
$2S_z$(1) as a function of $h_x$.
}
\label{v2sdwhx}
\end{figure}
\vskip 3mm

\begin{figure}
\mbox{\psfig{figure=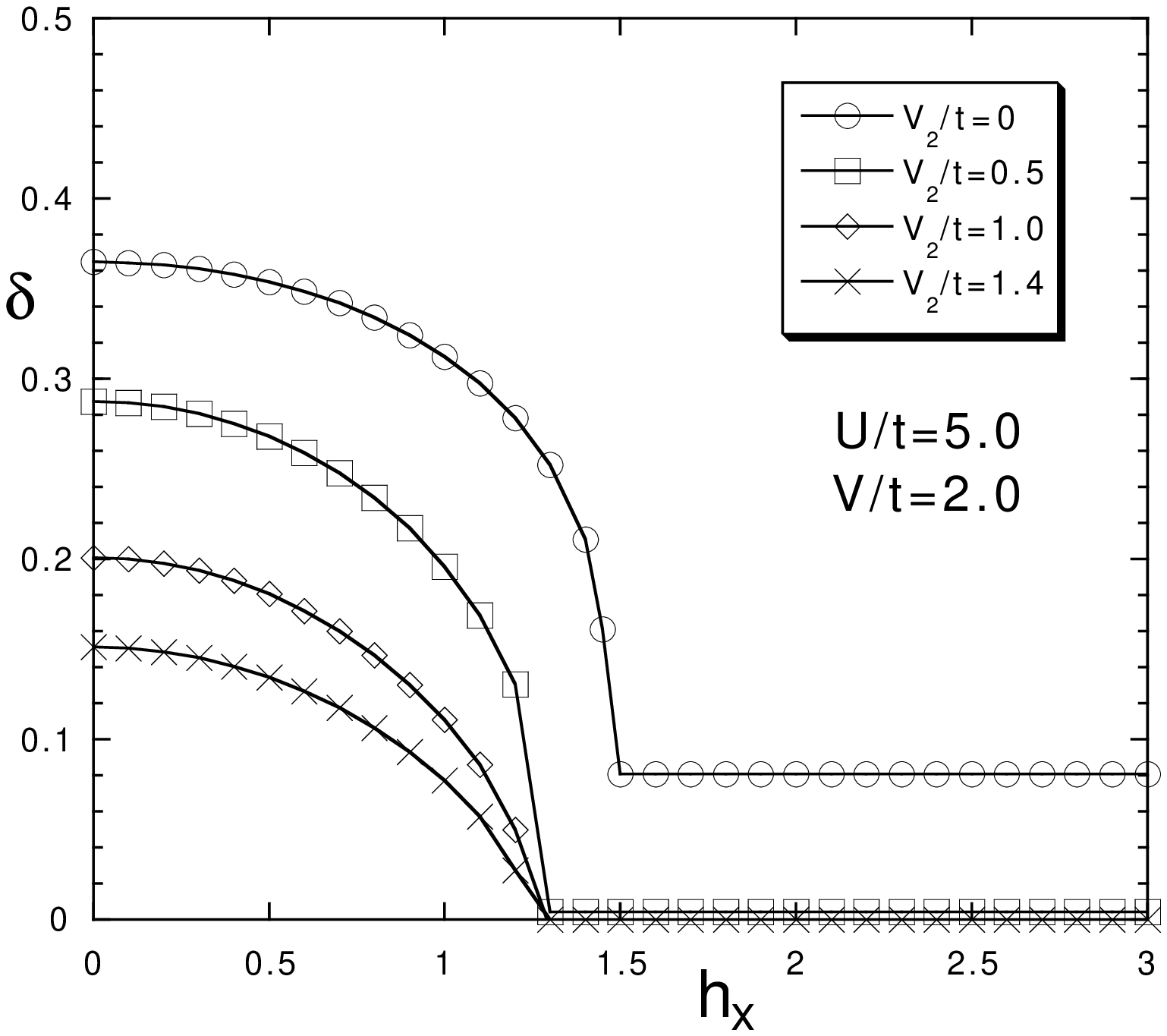,width=80mm,angle=0}}
\vskip 5mm
\caption{
$\delta$ as a function of $h_x$.
}
\label{v2cdwhx}
\end{figure}
\vskip 3mm

\begin{figure}
\mbox{\psfig{figure=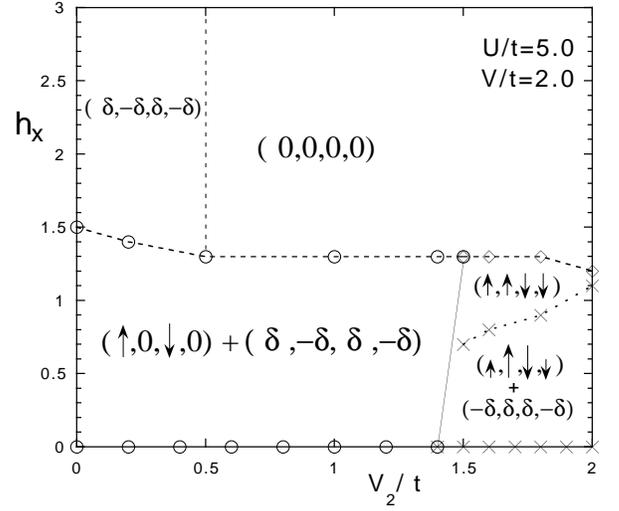,width=80mm,angle=0}}
\vskip 5mm
\caption{
$V-h_x$ phase diagram. 
}
\label{v2cdwhx}
\end{figure}
\vskip 3mm






\subsection{Finite Temperature
}

We show $S_z$ and $\delta$ as a function of $T$ at $H=0$. 
At $U/t=5.0$ and $V_2/t=0$, 
$S_z$ and $\delta$ as a function of $T$ for various values of 
$V$ are shown in Figs 
13 and 14, where 
$S_z(1,T)=S_z(2,T)=-S_z(3,T)=-S_z(4,T)$ for $V/t=0$ and 
$S_z(1,T)=-S_z(3,T), S_z(2,T)=S_z(4,T)=0$, 
$\delta(1,T)=\delta(3,T)=\delta(T)$ and 
$\delta(2,T)=\delta(4,T)=-\delta(T)$ for $V/t=0.5\sim5.0$. 
As $V$ increases, the critical temperatures ($T_{\rm SDW}$ 
and $T_{\rm CDW}$) at which 
$S_z$ and $\delta$ become zero increase. 
The $T$-dependences of $n$ for $V/t=3.0$, 4.0 and 5.0 have 
a dip at $T_{\rm SDW}$ for $V/t=3.0$, 4.0 and 5.0, 
which can be seen in Fig. 14.
We show the $V$-$T$ phase diagram in Fig. 15, 
where the solid and dotted lines are 
for the first and the second order transitions, respectively. 
Note that $T_{\rm CDW}>T_{\rm SDW}$ for $V/t>2.0$, 
that is, the charge order remains even if 
the spin order disappears under high 
temperatures. 

As we have shown above, 
the coexistent state of 
$2k_{\rm F}$-SDW and $2k_{\rm F}$-CDW 
(($_{\uparrow}$,$\uparrow$,$\downarrow$,$_{\downarrow}$) and 
($-\delta$,$\delta$,$\delta$,$-\delta$)) are 
stabilized for $V_2/t> 1.4$ at $T/t=0$ (see Figs. 6 and 7). 
At $T/t\neq0$, for example, for 
$V_2/t= 1.6$ as $T$ increases, 
the magnitudes of $S_z$ and $n$ decreases and 
the second order transition to the state of 
$2k_{\rm F}$-SDW
((${\uparrow}$,$\uparrow$,$\downarrow$,${\downarrow}$)) 
occurs, as shown in Figs. 16 and 17. 
Namely, $T_{\rm CDW}\leq T_{\rm SDW}$ in the coexistent state of 
$2k_{\rm F}$-SDW and $2k_{\rm F}$-CDW. 
For $V_2/t< 1.4$, 
$T_{\rm CDW}$ in the coexistent state of $2k_{\rm F}$-SDW and 
$4k_{\rm F}$-CDW is the same as $T_{\rm SDW}$, 
as shown in Figs. 18 and 19, where 
$S_z(1,T)=-S_z(3,T), S_z(2,T)=S_z(4,T)=0$, 
$\delta(1,T)=\delta(3,T)=\delta(T)$ 
and $\delta(2,T)=\delta(4,T)=-\delta(T)$. 
We show the $V_2$-$T$ phase diagram in Fig. 20,
where the solid and dotted lines are for the first and the second 
order transition lines, respectively. 

Our phase diagrams ($V$-$T$ and $V_2$-$T$) are 
in good agreement with those obtained by 
Tomio and Suzumura.\cite{tomio2}

\begin{figure}
\mbox{\psfig{figure=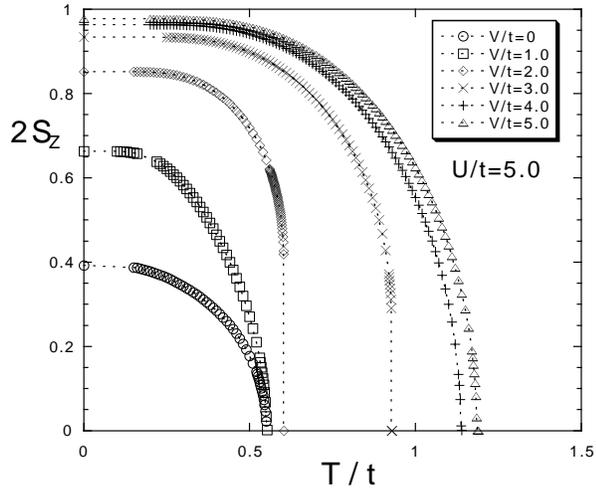,width=80mm,angle=0}}
\vskip 5mm
\caption{
2$S_z(1,T)$ as a function of $T/t$.
}
\label{v-tsz}
\end{figure}
\vskip 3mm

\begin{figure}
\mbox{\psfig{figure=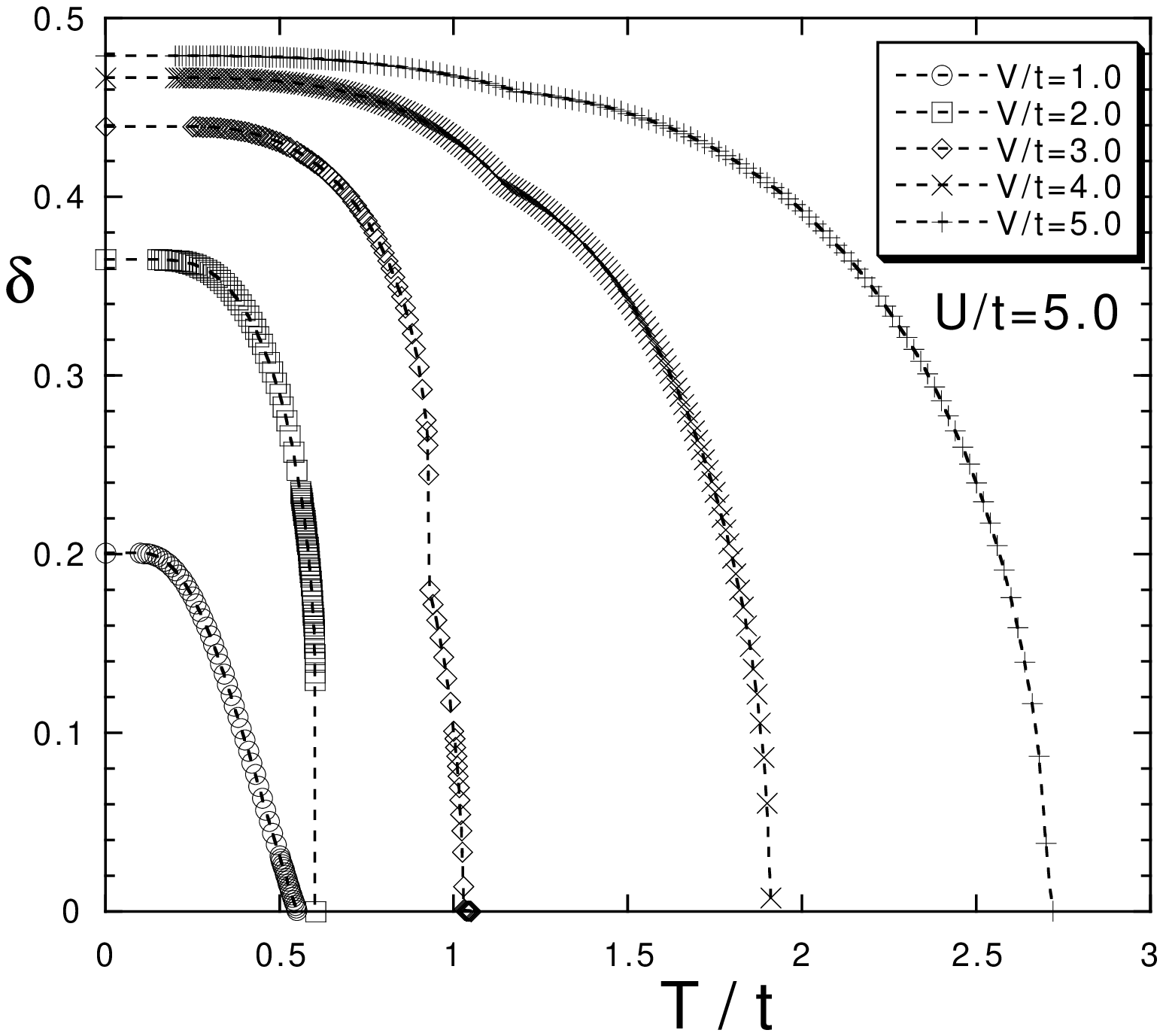,width=80mm,angle=0}}
\vskip 5mm
\caption{
$\delta(T)$ as a function of $T/t$.
}
\label{v-td}
\end{figure}
\vskip 3mm

\begin{figure}
\mbox{\psfig{figure=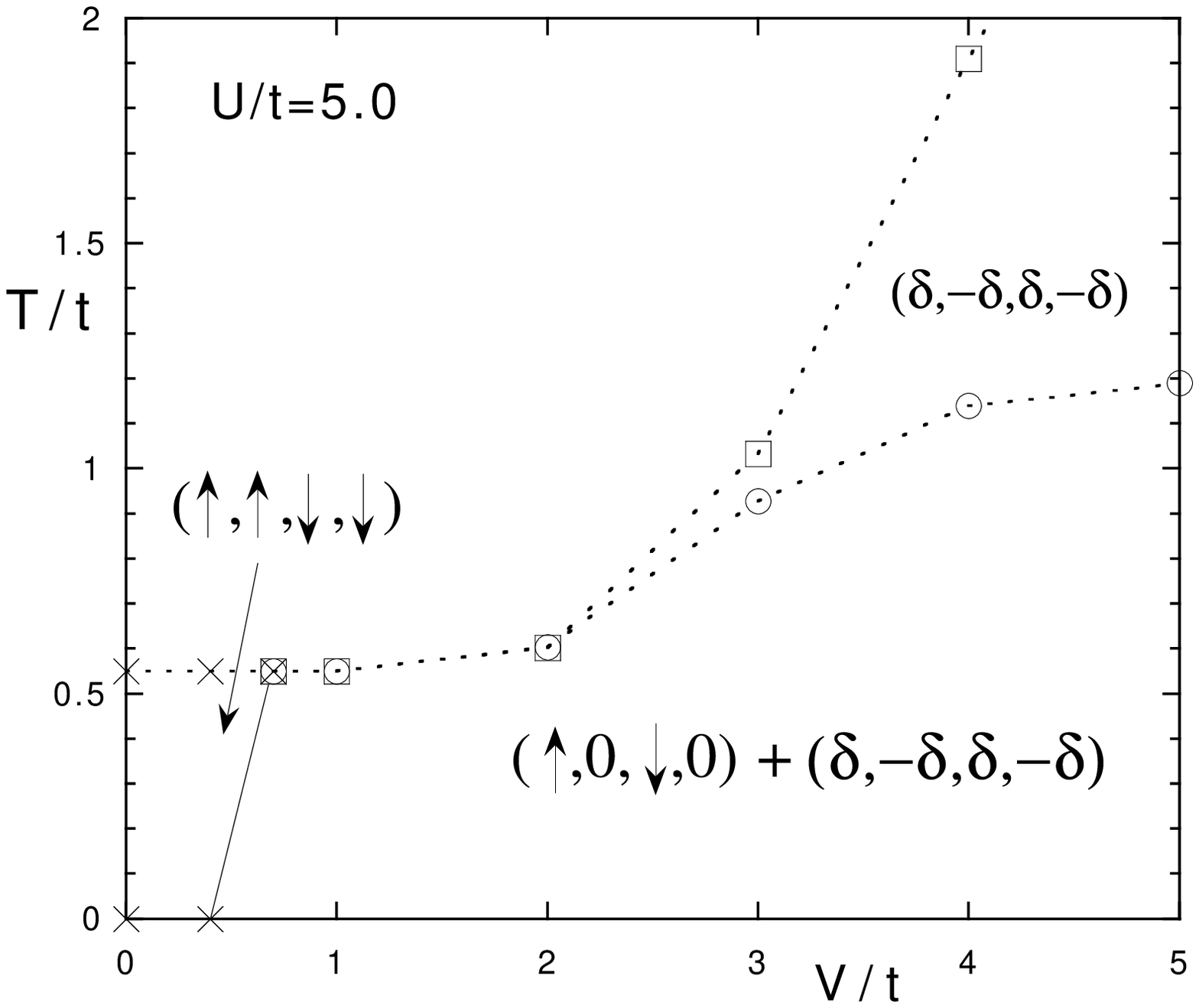,width=80mm,angle=0}}
\vskip 5mm
\caption{
$V-T$ Phase diagram.
}
\label{v-td}
\end{figure}
\vskip 3mm

\begin{figure}
\mbox{\psfig{figure=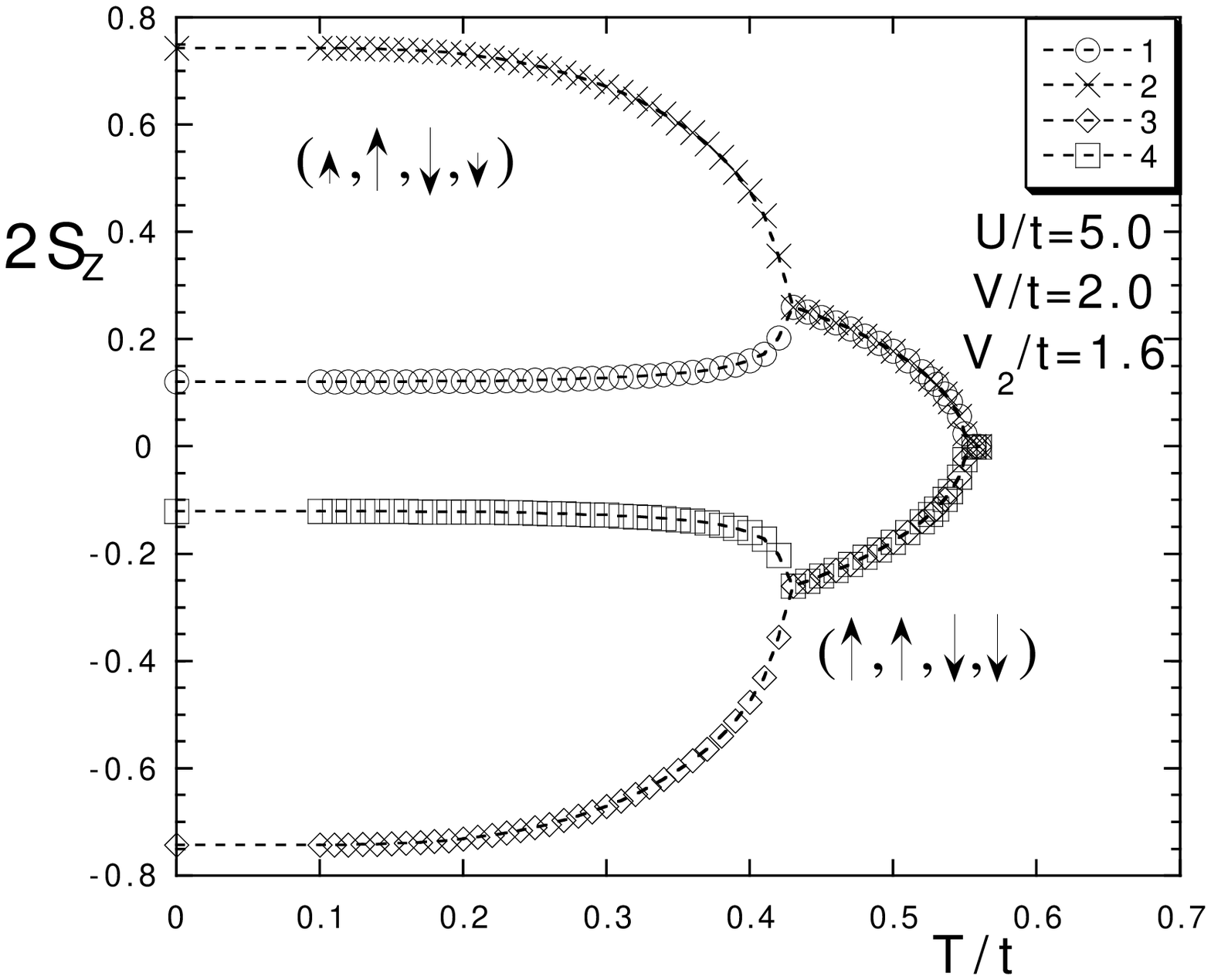,width=80mm,angle=0}}
\vskip 5mm
\caption{
2$S_z(j,T)$ $(j=1\cdots 4)$ as a function of $T/t$.
}
\label{vv16-t}
\end{figure}
\vskip 3mm

\begin{figure}
\mbox{\psfig{figure=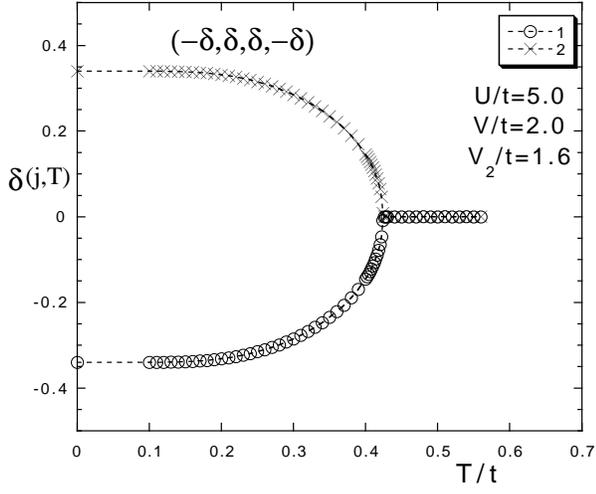,width=80mm,angle=0}}
\vskip 5mm
\caption{
$\delta(1,T)$ and $\delta(2,T)$ as a function of $T/t$.
}
\label{vv16-tcdw}
\end{figure}
\vskip 3mm

\begin{figure}
\mbox{\psfig{figure=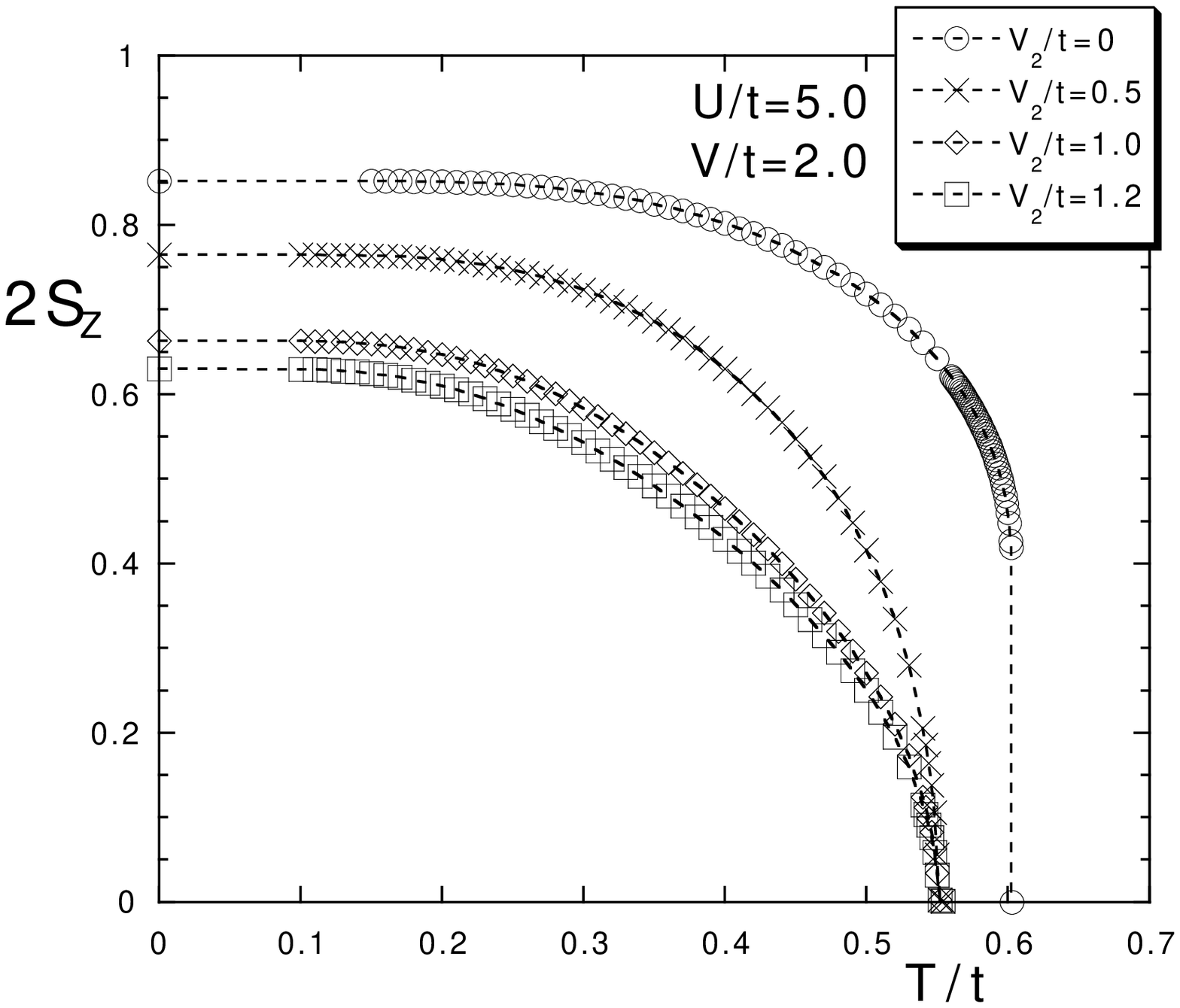,width=80mm,angle=0}}
\vskip 5mm
\caption{
2$S_z(1,T)$ as a function of $T/t$.
}
\label{v2sdw}
\end{figure}
\vskip 3mm

\begin{figure}
\mbox{\psfig{figure=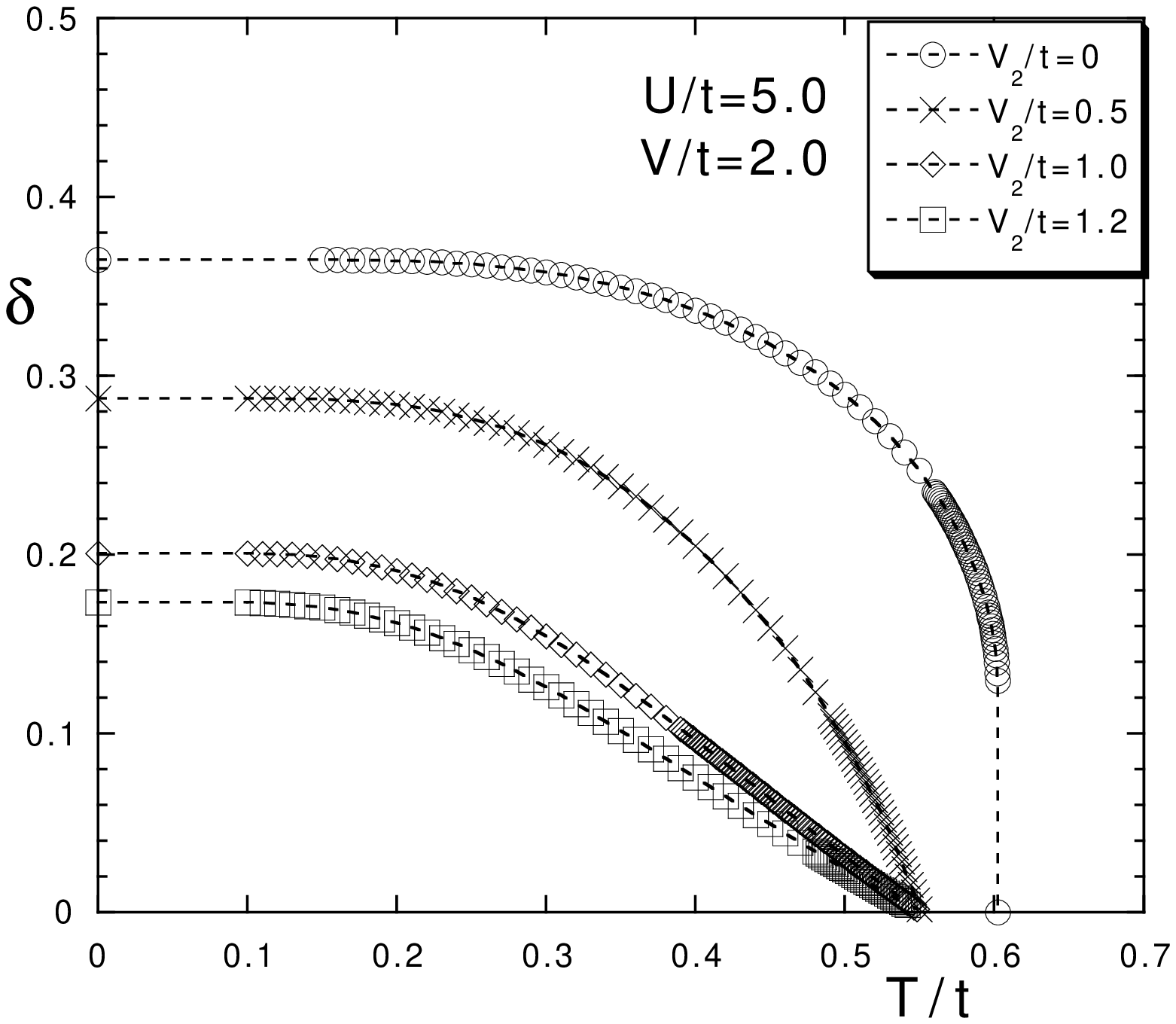,width=80mm,angle=0}}
\vskip 5mm
\caption{
$\delta(T)$ as a function of $T/t$.
}
\label{v2cdw}
\end{figure}
\vskip 3mm

\begin{figure}
\mbox{\psfig{figure=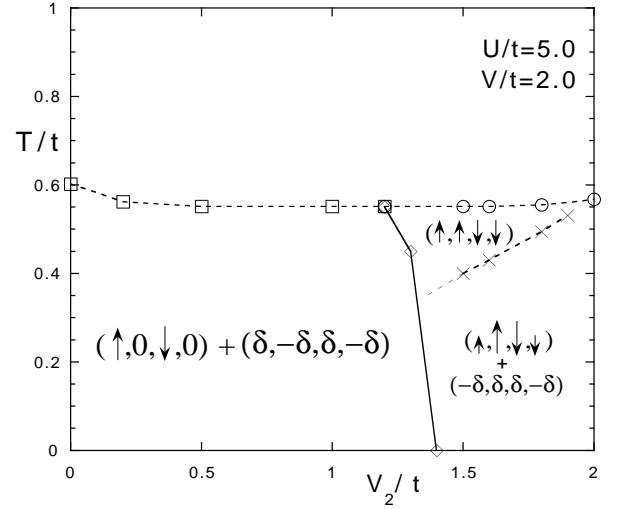,width=80mm,angle=0}}
\vskip 5mm
\caption{
$V_2-T$ Phase diagram.
}
\label{v-td}
\end{figure}
\vskip 3mm

\section{Conclusions}

We have shown the different 
dependences of the magnetic field and temperatures on 
the 2$k_{\rm F}$-SDW-2$k_{\rm F}$-CDW 
coexistent state and 
2$k_{\rm F}$-SDW-4$k_{\rm F}$-CDW 
coexisting state.

When $V_2$ is absent or small, 
the coexistent state of 
2$k_{\rm F}$-SDW and 4$k_{\rm F}$-CDW is stable. 
Then the charge order survives at 
high fields in the case 
of the perpendicular magnetic field to the easy axis when 
$V/t>2.0$ . 
This situation may be realized in 
(TMTTF)$_2$$X$, where 4$k_{\rm F}$-CDW is expected to be observed 
in X-ray scattering measurements 
under high fields.

On the other hand, 
if $V_2$ is large enough, the coexistent state of 
2$k_{\rm F}$-SDW and 2$k_{\rm F}$-CDW is stabilized. 
When the magnetic field is applied perpendicular 
along the easy axis, 
the coexistent state of 
2$k_{\rm F}$-SDW and 2$k_{\rm F}$-CDW changes 
to 2$k_{\rm F}$-SDW state. 
This transition is the second order phase transition, 
and 
($_{\uparrow}$,$\uparrow$,$\downarrow$,$_{\downarrow}$) 
becomes 
($\uparrow$,$\uparrow$,$\downarrow$,$\downarrow$). 
This transition may be observed in the 
angle dependence of satelite peak positions of 
NMR in (TMTSF)$_2$PF$_6$. 

The critial temperature of the charge 
order ($T_{\rm CDW}$) is higher than that of 
the spin order ($T_{\rm SDW}$) in the coexistent phase 
of 2$k_{\rm F}$-SDW and 4$k_{\rm F}$-CDW when 
$V/t>2.0$. This is in good agreement with 
$T_{\rm SDW}\sim 13$ K and $T_{\rm CDW}\sim 100$ K 
observed in X-ray scattering in 
(TMTTF)$_2$Br. 
Since 
$T_{\rm CDW}$ is convergent to $T_{\rm SDW}$ for $V/t<2.0$,
we expect that 
the critical temperatures of 2$k_{\rm F}$-SDW and 4$k_{\rm F}$-CDW 
will become the same 
when 
the pressure is applied to (TMTTF)$_2$Br.

In the coexistent phase of 2$k_{\rm F}$-SDW and 2$k_{\rm F}$-CDW, 
we find that $T_{\rm CDW}\leq T_{\rm SDW}$, which is consistent with the X-ray 
scattering measurement in (TMTSF)$_2$PF$_6$, 
where $T_{\rm CDW}\simeq T_{\rm SDW}$.\cite{pouget}


\section{Acknowledgment}
The authors would like to thank
T. Sakai, K. Machida, Y. Nogami, Y. Suzumura and Y. Tomio 
for valuable discussions.

\widetext

\begin{references}
\bibitem{review}
For a review, see: T. Ishiguro, K. Yamaji, and G. Saito: {\it
Organic
Superconductors}
(Springer-Verlag, Berlin  1998).

\bibitem{jerome}
For a review, see D. Jerome: {\it
Organic
Conductors} ed J. P. Farges
(Marcel Deckker, New York, 1994).

\bibitem{takahashi}
T. Takahashi, Y. Maniwa, H. Kawamura and
G. Saito: J. Phys. Soc. Jpn. {\bf 55}
(1986) 1364.

\bibitem{delrieu}
J. M. Delrieu, M. Roger, Z. Toffano and A. Moradpour:
J. Phys. (Paris) {\bf 47}
(1986) 839.


\bibitem{pouget}
J. P. Pouget and S. Ravy:
Synth. Met. {\bf 85} (1997) 1523.


\bibitem{barthel}
E. Barthel, G. Quirion, P. Wzietek, D. Jerome, J. B. Christensen,
M. Joregensen and K. Bechgaard:
Europhys. Lett. {\bf 21}
(1993) 87.

\bibitem{nakamura}
T. Nakamura, T. Nobutoki, Y. Kobayashi, T. Takahashi and
G. Saito: Synth. Met. {\bf 70} (1995) 1293.



\bibitem{nakamura2}
T. Nakamura, R. Kinami, Takahashi and
G. Saito: Synth. Met. {\bf 86} (1997) 2053.


\bibitem{seofukuyama}
H.Seo and H. Fukuyama: J. Phys. Soc. Jpn. {\bf 66}
(1997) 1249.

\bibitem{nobuko}
N. Kobayashi and M. Ogata: J. Phys. Soc. Jpn. {\bf 66}
(1997) 3356.


\bibitem{nobuko2}
N. Kobayashi, M. Ogata and K. Yonemitsu:
J. Phys. Soc. Jpn.
{\bf 67} (1998) 1098.


\bibitem{Mazumdar}
S. Mazumdar, S. Rammasesha, R. Torsten Clay and
David K. Campbell:
Phys. Rev. Lett. {\bf 82} (1999) 1522.

\bibitem{yoshioka}
H. Yoshioka, M. Tsuchiizu, and Y. Suzumura, 
cond-mat/0001450

\bibitem{mila}
F. Mila:
Phys. Rev. B {\bf 52} (1995) 4788.

\bibitem{pauli}
B. S. Shandrasekhar: Appl. Phys. Lett. {\bf 1} (1962) 7.


\bibitem{pauli2}
A. M. Clogston: Phys. Rev. Lett. {\bf 9} (1962) 266.

\bibitem{Mckenzie}
For study of the Pauli limit of CDW in recent,
Ross H. McKenzie cond-mat/9706235.

\bibitem{kishigi}
K. Kishigi and Y. Hasegawa: cond-mat/0002165



\bibitem{band}
T. Mori, A. Kobayashi, Y. Sasaki and H. Kobayashi:
Chem. Lett. (1982) 1923.

\bibitem{band2}
P. M. Grant: J. Phys. Colloq. C{\bf 3} (1983) 847.

\bibitem{band3}
T. Mori, A. Kobayashi, Y. Sasaki and H. Kobayashi, G. Saito and
H. Inokuchi: Bull.
Chem. Soc. Jpn. {\bf 57} (1984) 627.

\bibitem{band4}
L. Ducasse, M. Abderrabba, J. Hoarau, M. Pesquer,
B. Gallois and J. Gaultier: J. Phys. C{\bf 19} (1986) 3805.

\bibitem{ising}
B. Bleaney: Proc. R. Soc. London, Ser A {\bf 276} (1963) 19.


\bibitem{ising2}
V. Yu. Irkhin and A. A. Katanin: Phys. Rev. B{\bf 58} (1998) 5509.

\bibitem{spin}
For example, for (TMTSF)$_2$AsF$_6$, 
K. Mortensen, Y. Tomkiewicz, K. Bechgaard: Phys. Rev. {\bf 25} 
(1982) 3319.

\bibitem{fulde}
P. Fulde and A. Ferrel: 
Phys. Rev. {\bf 135} (1964) A550. 

\bibitem{fulde2}
A. I. Larkin and Yu. N. Ovchinnikov: 
Sov. Phys. JETP {\bf 20} (1965) 762.

\bibitem{tomio}
Y. Tomio and Y. Suzumura,
J. Phys. Soc. Jpn.
{\bf 69} (2000) 796.

\bibitem{tomio2}
Y. Tomio and Y. Suzumura,
J. Phys. Soc. Jpn.
preprint.




\end{references}
\end{document}